\documentclass[superscriptaddress,twocolumn,prb,showpacs]{revtex4-1}

\usepackage{graphicx}
\usepackage{subfigure}

\usepackage{amsmath}
\usepackage{amssymb}
\usepackage{amsthm}
\usepackage{hyperref}

\newcommand{\Tr}{\mathrm{Tr}}

\begin{document}
\title[]{Singles correlation energy contributions in solids}

\author{Ji\v{r}\'{i} Klime\v{s}}
\affiliation{Department of Chemical Physics and Optics, Faculty of Mathematics and Physics, 
Charles University, Ke Karlovu 3, CZ-12116 Prague 2, Czech Republic}
\affiliation{J. Heyrovsk\'{y} Institute of Physical Chemistry, Academy of Sciences of the Czech Republic, 
Dolej\v{s}kova 3, CZ-18223 Prague 8, Czech Republic}

\author{Merzuk~Kaltak}
\affiliation{University of Vienna, Faculty of Physics and Center for 
Computational Materials Science, Sensengasse 8/12, A-1090 Vienna, Austria}

\author{Emanuele Maggio}
\affiliation{University of Vienna, Faculty of Physics and Center for 
Computational Materials Science, Sensengasse 8/12, A-1090 Vienna, Austria}

\author{Georg~Kresse}
\email{georg.kresse@univie.ac.at}
\affiliation{University of Vienna, Faculty of Physics and Center for 
Computational Materials Science, Sensengasse 8/12, A-1090 Vienna, Austria}

\date{\today }

\begin{abstract} 
The random phase approximation to the correlation energy often yields 
highly accurate results for condensed matter systems.
However, ways how to improve its accuracy are being sought
and here we explore the relevance of singles contributions for prototypical
solid state systems. We set out with a derivation of the random phase
approximation using the adiabatic connection and fluctuation dissipation theorem, but
contrary to the most commonly used derivation, the  density is allowed to vary along
the coupling constant integral. This yields results closely paralleling
standard perturbation theory. We  re-derive
the standard singles of G\"orling-Levy perturbation theory 
[G\"orling and Levy, Phys. Rev. A {\bf 50}, 196 (1994)],
highlight the analogy of our expression to the renormalized
singles introduced by Ren and coworkers 
[Ren, Tkatchenko, Rinke, and Scheffler, Phys. Rev. Lett. {\bf 106}, 153003 (2011)],
and introduce a new approximation for the singles using the density
matrix in the random phase approximation. 
We discuss the physical relevance and importance of singles alongside illustrative examples of
simple weakly bonded systems, including rare gas solids (Ne, Ar, Xe), ice, adsorption of
water on NaCl, and solid benzene.
The effect of singles on covalently and metallically bonded systems is also discussed.
\end{abstract}

\maketitle

\section{Introduction}

In the last decade the interest in many body perturbation theory has 
risen significantly. This is to some extent related to the enormous
increase in the available computer performance, but it is also
driven by the realization that many of the presently available density
functionals have limited predictive accuracy. Improving density functionals
is a very active field of research in itself.  In fact, tremendous progress
has been made for semiconductor and insulator modelling by
the inclusion of exact exchange,\cite{Becke.98.1372,Becke.98.5648,Muscat2001397,Paier.122.234102,Paier.124.154709}  
as well as for weakly bonded systems by including either
atom centered dispersion
corrections \cite{grimme2006d2,tkatchenko2009,tkatchenko2012,grimme2011} or non-local
van der Waals corrections regarding the density at two points in
space.\cite{dion2004,langreth2005,langreth2009,vydrov2010VV10,klimes2012}
However, a unified comprehensive, accurate and predictive framework for metallic, covalently as well as 
dispersion driven interactions is hard to attain using present
density functionals: most available density 
functionals require careful evaluation against more accurate methods
before one can trust them to predict accurate numbers for
a specific material.

In the quantum chemistry community, such a concise hierarchy of
methods for evaluating and benchmarking more approximate methods, such
as density functional theory, is well established. The highest rung of
this hierarchy is made up by the full configuration interaction method, followed by 
a variety of more approximate methods. For instance, 
if the material is dominated by a single Slater determinant,
the methods of choice are coupled cluster methods,\cite{Bartlett_RevModPhys_2007}
as well as  M{\o}ller-Plesset perturbation theory\cite{Moeller-Plesset_PhysRev_1934} for large band gap systems.
Only recently these methods have become available for solids.\cite{Marsman_MP2_2009,Gruneis_MP2_2010,Booth_FCIQM_nature_2013}

For solids, calculations using coupled cluster methods or full configuration interaction
methods are, however, exceedingly demanding approaching several 100.000 CPU hours for
a single material with a few atoms in the unit cell. 
The approach taken in solid state systems is therefore
often ``bottom up'', {\em i.e.} starting with an approximate scheme such as
density functional theory and improving
upon the description until results compare reasonably well with experiment. 
The random phase approximation (RPA) is one promising approach 
to achieve this goal.\cite{Langreth_ACFFDT_1977}
In fact, the RPA  yields a balanced description of most
bonding types, including metallic bonding, covalent, ionic, 
as well as van der Waals (vdW) bonding.
Initial applications were limited to small molecules.
\cite{Furche_RPA_mol_2001} 
Although first applications to bulk materials were not encouraging,\cite{Miyake_RPA_2002}
for many materials results nowadays surpass those from  semi-local
functionals.\cite{Marini_BN_2006,harl2008,Furche_RPA_2008,Eshuis_RI_RPA_2010}
The studies now span a wide range of applications,
including molecular reactions,\cite{Eshuis_RPA_works_2011}
rare gas solids,\cite{harl2008} properties of covalent, metallic
and ionic solids,\cite{harl2009,harl2010,Xiao_RPA_Si_2012,Olsen_RPA_solids_mol_2013,schimka2013} dispersion forces in graphite and 
between graphene and surfaces,\cite{lebegue2010,Olsen_graphene_2011,Mittendorfer_graphene_Ni_2011}
layered compounds,\cite{Gulans_RPA_layered_2012}
adsorption of molecules on surfaces,\cite{Schimka2010} bulk ice properties,\cite{Macher} and many more applications are emerging.
Recent advances include highly efficient implementations scaling only cubically with system size
and linearly in the number of k-points,\cite{Kaltak_LowScaling_2014,Kaltak_SiLowScaling_2014}
as well as implementations scalable to massively parallel computers.\cite{DelBen_RPAparallel_2015}

There is no denying that RPA is not perfect. Among all the possible
many body diagrams, direct RPA exclusively sums
the bubble diagrams. Attempts to include other kinds of diagrams, for instance higher order
exchange interactions,\cite{Grueneis_SOSEX_2009,furche2013} the contribution
of single excitations \cite{ren2011,Ren_Singles_2012,paier2012,ren2012,ren_renormalized_2013}  
or approximate 
exchange and correlation contributions inspired  by density functional theory \cite{Olsen_fxc_ALDA_2012}  are currently vigorously explored research directions.
Also better starting points than standard density functionals are
explored.\cite{bleiziffer2013,klimes2014}
Likewise, forces are yet only  implemented in two molecular
codes,\cite{Burow_RPAforces_2014,Helgaker_RPAforces_2013} and
they are not available in solid state codes.

The present paper mainly concentrates on the already mentioned singles,
contributions that arise from determinants where one electron is moved
from a state occupied in the initial Slater determinant to an originally unoccupied state. 
If the initial determinant is the Hartree-Fock determinant, then 
the singles are zero in lowest
order perturbation theory  because of Brillouin's theorem.\cite{Szabo_QC_1996}
If one sets out from the Kohn-Sham determinant, the singles however
contribute even in lowest order perturbation theory. 
In the adiabatic-connection framework this contribution was
first derived by G\"orling and Levy.\cite{Goerling_Levy_GL2_1994}
In the present work, for reasons of consistency we will re-derive
the same term, albeit doing so using a concise Green's function notation, 
making the derivation algebraically simpler (see Sec. \ref{sec:theory_GL2}). 
The main point of our work is, however, 
that we give up the assumption that the charge density is  kept constant along the coupling constant integration 
(see Sec. \ref{sec:theory_ACFDT}), an approximation  made in most derivations of 
the adiabatic-connection fluctuation-dissipation theorem 
(AC-FDT).\cite{Langreth_ACFFDT_1977} 
This approximation was originally adopted, 
since density functionals were considered to be very accurate in predicting the groundstate density, 
and because AC-FDT was used as a theoretical pathway to derive improved density functionals. 
However, since AC-FDT is
now a computational framework, and since density functionals do
not always yield accurate densities, we feel that this point urgently
needs to be revised. We are aware of at least two papers
where this assumption has been  given up as well, albeit in the first one the point
was only made in passing\cite{Toulouse_range-separated_2010} and
the second one only considers small prototypical molecules.\cite{Aggelen_AC_2014}

Our present formal derivation (see Sec. \ref{sec:theory_singles}) yields results similar
to the singles and re-normalized singles proposed by Ren et al.\cite{ren2011,Ren_Singles_2012,ren_renormalized_2013}
Ren and coworkers, however, formally gave up the adiabatic-connection framework
and used standard second order perturbation theory to motivate
their singles and re-normalized singles. 
We feel that a concise perturbational framework (here the adiabatic-connection framework) 
is helpful to better understand the underlying approximations.
We will discuss that the singles account for the changes of the mean
field density matrix along the coupling constant integral. 
This is exactly analogous to coupled cluster theory or, since
coupled cluster theory is just a re-summation of certain Goldstone diagrams, 
standard many body perturbation theory.
This insight explains why the inclusion of singles 
increases the bonding between weakly interacting fragments, i.e. atoms
and molecules. When the singles are included the charge density contracts compared to
the original density functional, and this results in a decrease of the Pauli repulsion.
We demonstrate this effect here for rare gas solids, as
well as, the cubic phase of ice, the benzene crystal and water adsorption on NaCl. 
For these  systems, the predicted cohesive energies and lattice constants
are, after inclusion of the singles, in very good agreement with experiment 
(see Sec. \ref{sec:raregas}, \ref{sec:ice}, \ref{sec:benzene}, \ref{sec:water}).
We also present results for covalently bonded as well as metallic systems.
Here no improvements are discernible, or rather, the corrections
of the singles are mostly tiny and do not worsen
the already excellent agreement with experiment (see Sec. \ref{sec:solids}).

\section{Theory}
\label{sec:theory}

\subsection{Adiabatic switching with density change}
\label{sec:theory_ACFDT}

In this section, we give a  brief derivation of the adiabatic switching, 
where the correlation energy is obtained by switching from the single determinant reference
system to the many body system of interest.
We consider the usual adiabatic connection, where the density is kept constant
along the pathway, as well as adiabatic switching allowing for a change of the density.

In the adiabatic-connection framework, it is common practice to 
switch from the purely local, in real space multiplicative potential $\hat V_\lambda$ to the exact many electron Hamiltonian $\hat{H}_{\lambda=1}$:
\begin{equation}
\label{equ:H}
 \hat{H}_\lambda= \sum_i \hat{h}({\bf r}_i) + \frac{\lambda}{2}\sum_{i\ne j} \hat{v}(\mathbf{r}_i,\mathbf{r}'_j) + \sum_i \hat{V}_\lambda({\bf r}_i).
\end{equation}
Here $\hat{v}$ is a two-electron operator, typically the Coulomb kernel $1/| {\bf r}- {\bf r}'|$, and
$\hat{h}$ is an arbitrary one electron operator, typically the kinetic
energy operator $\Delta$ and the potential of the ionic cores or some other external one-electron 
potential $\hat{V}^{\rm ext}$
\[
\label{equ:h}
  \hat{h}({\bf r}) = -\frac{1}{2} \Delta_{\bf r}  + \hat{V}^{\rm ext}({\bf r}),
\] 
where we have assumed atomic (Hartree) units.
The correlation energy $E^{\rm tot}_c$ is defined as the difference between a single
Slater determinant $\Psi_0$ evaluated using the orbitals at zero
coupling ($\lambda=0$) and the exact many-electron wave function  $\Psi_1$
evaluated at full coupling ($\lambda=1$):
\begin{equation}
\label{equ:Ec}
 E^{\rm tot}_c = \langle \Psi_1 | \hat{H}_1 |  \Psi_1 \rangle - \langle \Psi_0 | \hat{H}_1 |  \Psi_0 \rangle.
\end{equation}
At zero coupling, the  Hamiltonian is chosen to be the usual Kohn-Sham Hamiltonian,
\begin{equation}
 \left(-\frac{1}{2} \Delta  + \hat{V}^{\rm ext} +  \hat{V}^{\rm KS}_0\right) | \phi_n \rangle = \epsilon_n | \phi_n \rangle,
\end{equation}
where $\hat{V}^{\rm KS}_0$ is the standard Kohn-Sham potential (we keep the subscript $0$
to indicate zero coupling). 
Note that the Kohn-Sham potential includes exchange, correlation, and the Hartree contributions.
In the following, we will also use the shorthand $| n \rangle = | \phi_n \rangle$, and the indices
$i,j,k, ...$ always refer to occupied one-electron orbitals, whereas the
indices $a,b,c,...$ refer to virtual (unoccupied) one-electron orbitals of the Kohn-Sham
Hamiltonian.

Subtracting and adding $\hat{V}^{\rm KS}_0$ to the second term on the r.h.s of (\ref{equ:Ec}) one can
rewrite the correlation energy as 
\begin{equation}
\begin{split}
 E^{\rm tot}_c & = \langle \Psi_1 | \hat{H}_1 |  \Psi_1 \rangle - \langle \Psi_0 | \hat{H}_0 |  \Psi_0 \rangle \\
     & -  \frac{1}{2}\langle \Psi_0 | \hat{v} |  \Psi_0 \rangle + \langle \Psi_0 |\hat{V}^{\rm KS}_0 |  \Psi_0 \rangle  \\
 E^{\rm tot}_c & = \int_0^1 \langle \Psi_\lambda | \frac{d\, \hat{H}_\lambda}{d\, \lambda} |  \Psi_\lambda \rangle d \lambda  \\
     & - \frac{1}{2} \langle \Psi_0 | \hat{v} |  \Psi_0 \rangle + \langle \Psi_0 |\hat{V}^{\rm KS}_0 |  \Psi_0 \rangle .
\end{split}
\end{equation}
In going from the first to the second equation, the ``Hellman-Feynman'' theorem has been used, {\em i.e.}
it is assumed that  $\Psi_\lambda$ is the groundstate wave function
of the Hamiltonian $\hat{H}_\lambda$, and terms involving the derivative of
the wave function $d\, \Psi_\lambda / d\, \lambda $ thus vanish.
The first term in the second line is exactly the Hartree and exchange energy evaluated for
Kohn-Sham orbitals. Inserting the derivative of the Hamiltonian $d\, H_\lambda / d\, \lambda $
immediately yields (compare Equ. \eqref{equ:H})
\begin{equation}
\begin{split}
\label{eq:ccintegral}
 E^{\rm tot}_c & = \frac{1}{2}\int_0^1 \langle \Psi_\lambda |  \hat{v} |  \Psi_\lambda \rangle d \lambda - \frac{1}{2} \langle \Psi_0 | \hat{v} |  \Psi_0 \rangle \;  +\\
     &  \int_0^1 \langle \Psi_\lambda | \frac{d\, \hat{V}_\lambda}{d\, \lambda} |  \Psi_\lambda \rangle d \lambda+ \langle \Psi_0 |\hat{V}^{\rm KS}_0 |  \Psi_0 \rangle\,. 
\end{split}
\end{equation}
We note that a similar term is given in the appendix of 
Ref.~\onlinecite{Toulouse_range-separated_2010} without a clear derivation.
The most common way  to realize the coupling integral is to chose the potential $\hat V_\lambda$ such
that the density remains constant along the coupling constant integral.\cite{Langreth_ACFFDT_1977}
Then the expectation value of the density operator yields the (constant) groundstate density
$\langle \Psi_\lambda | \hat n({\bf r}) |  \Psi_\lambda \rangle= n({\bf r}) $, 
and the first term in the second line can be integrated yielding
\[
\int_0^1 \langle\Psi_\lambda| \frac{d\, \hat{V}_\lambda}{d\,
\lambda}|\Psi_\lambda\rangle d \lambda = \langle\Psi_0| 
\hat{V}_1-\hat{V}^{\rm KS}_0\ |\Psi_0\rangle\,,
\]
canceling the last term in the second line of Equ. (\ref{eq:ccintegral}). 
Since the additive potential $\hat V_\lambda$ must be  zero at full coupling $\hat{V}_1=0$, one obtains
the standard equation for the AC-FDT:
\begin{equation}
\label{eq:ccintegral_rhoc}
E_c = \frac{1}{2}\int_0^1 \langle \Psi_\lambda |  \hat{v} |  \Psi_\lambda \rangle d \lambda  -\frac{1}{2} \langle \Psi_0 | \hat{v} |  \Psi_0 \rangle\,,
\end{equation}
which is only valid {\em if the density is constant along the integration pathway}, whereas 
the full version is obviously given by Equ. (\ref{eq:ccintegral}).

To obtain a simplified version of the full equation, one can 
switch off  the Kohn-Sham potential {\em linearly} {\em i.e.}
\begin{equation}
\label{equ:linear}
 \hat{V}_\lambda = (1-\lambda) \hat{V}^{\rm KS}_0\,.
\end{equation} 
Linear switching was first considered by Harris and Jones,\cite{Harris_AC_1974}
 and subsequently discussed in Refs.~\onlinecite{Toulouse_range-separated_2010}
and~\onlinecite{Aggelen_AC_2014}. Linear switching is also exact (as long
as no phase transition is encountered) and yields
results identical to standard Rayleigh-Schr\"odinger perturbation theory where the perturbation
is also switched on linearly. 

Linear switching yields for the second line in Equ.~(\ref{eq:ccintegral}) 
a simple correction to the correlation energy
\begin{equation}
\begin{split}
\label{equ:ccintegral_rho}
 \Delta E_c  &= - \int_0^1 d\lambda \int d^3 {\bf r}\, n_\lambda({\bf r}) \hat{V}^{\rm KS}_0({\bf r} )+ \int  d^3 {\bf r}\, n_0({\bf r}) \hat{V}^{\rm KS}_0({\bf r} ) \\
     & = - \int_0^1 d\lambda \int d^3 {\bf r}\, \left(n_\lambda({\bf r})-n_0({\bf r}) \right) \hat{V}^{\rm KS}_0({\bf r} ).
\end{split}
\end{equation}
Here $n_\lambda$ is the charge density at coupling $\lambda$.
The total correlation energy is then given by the sum of Equ.~(\ref{eq:ccintegral_rhoc})
and Equ.~(\ref{equ:ccintegral_rho}).

\subsection{Fluctuation-Dissipation theorem}
\label{sec:theory_FD}

It is common to rephrase the correlation energy in 
Equ.~(\ref{eq:ccintegral_rhoc}) using the fluctuation-dissipation theorem. 
The derivation is sketched in appendix A, however,
similar derivations can be found elsewhere.\cite{Langreth_ACFFDT_1977,Fuchs_RPA_2005,Toulouse_range-separated_2010,Aggelen_AC_2014} 
The final result becomes:
\begin{equation}
\label{equ:acfdt}
\begin{split}
 E_c & = \frac{1}{2} \int_0^1 \left([n_\lambda \hat v n_\lambda] -[n_0 \hat v
n_0]\right) \, d\lambda  \\
  &-\frac{1}{2}  \int_0^1   \Tr [(\chi_\lambda(0^-)+n_\lambda -\chi_0(0^-)-n_0) \hat v] d\lambda \,.  \\
 \end{split}
\end{equation}
Here we have introduced short-hands
\begin{equation}
\begin{split}
  [n \hat v n] &= \int  n({\bf r}) \hat v({\bf r},{\bf r}')  n({\bf r}') d^3{\bf r}d^3{\bf r}'\\
  \Tr [(\chi + n)\hat v] &=\int  (\chi({\bf r}, {\bf r}')+ n({\bf r}) \delta({\bf r}-{\bf r}') ) \hat v({\bf r}', {\bf r}) d^3 {\bf r} d^3 {\bf r}'.
\end{split}
\end{equation}
In the equations above,   $\chi_\lambda$ is the reducible polarizability 
(or density fluctuation response function)
of the many electron system at coupling $\lambda$:
\begin{equation}
\label{eq:chi_def}
\chi_\lambda(\tau, {\bf r}, {\bf r}') =  - \langle \psi_\lambda | \mathcal{T} [\delta \hat n(\tau,{\bf r}) \delta \hat n(0,{\bf r}')] | \psi_\lambda \rangle.
\end{equation}
$\mathcal{T}$ is the time ordering operator, and $\delta \hat n$ the density fluctuation operator 
$\delta \hat n = \hat n - \langle  \hat  n \rangle_\lambda$.
The polarizabilities are evaluated at a small negative infinitesimal imaginary time $\tau \to 0^-$.
Note that throughout the paper, $\tau$ is the imaginary part
of an imaginary time $t=-i\tau$. 
The polarizability in imaginary time can be related to the polarizability in 
imaginary frequency $i \omega$ by a Fourier transformation:
\begin{equation}
\begin{split}
\label{equ:ft}
  \chi_\lambda(\tau) &= \frac{1}{ 2 \pi}\int_{-\infty}^{\infty} d \omega \exp(-i \omega \tau) \bar \chi_\lambda(i \omega) \\
\bar \chi_\lambda(i \omega) &= \quad \;\;\int_{-\infty}^{\infty} d \tau \exp(i \omega \tau)   \chi_\lambda(\tau). 
\end{split}
\end{equation}
At zero coupling, $\chi_0$ becomes the independent particle polarizability $P_0$ of the Kohn-Sham system, and can be written in terms of the one particle Green's function\cite{Hedin_GW_1965,Toulouse_range-separated_2010}
We define the independent particle polarizability at coupling $\lambda$ generally as
\begin{equation}
 \label{equ:chi}
  P_\lambda(\tau,{\bf r}, {\bf r}')= G_\lambda(\tau,{\bf r}, {\bf r}')G_\lambda(-\tau, {\bf r}', {\bf r}),
\end{equation}
where  $G_\lambda$ is the
one-particle Green's function of the $\lambda$ interacting system $\hat H_\lambda$.
At zero coupling, {\em i.e.}  for the Kohn-Sham case, the Green's function in imaginary time $G_0$  are  given by ($\mu$ is the
chemical potential of the electrons, $i$ are occupied and $a$ are unoccupied one electron orbitals):
\begin{equation}
\label{equ:Greens}
G_0(\tau) = \left \{ \begin{array} {llr}
  &\sum_i |i \rangle \langle i|~ \exp(-(\epsilon_i-\mu) \tau)  & \tau<0 \\ 
 - &\sum_a |a \rangle \langle a| \exp(-(\epsilon_a-\mu) \tau)  & \tau>0 \\ 
\end{array} \right. .
\end{equation}

Before continuing, we note that  at small $\lambda$ and neglecting fluctuations, the term in the integral on the second line in Equ. (\ref{equ:acfdt}) can be also written as an integral of the exchange energy
\begin{equation}
\label{equ:acfdt_ex}
 \approx  - \frac{1}{2}  \int_0^1   \gamma_\lambda( {\bf r},{\bf r}')\gamma_\lambda({\bf r}',{\bf r}) \hat v({\bf r}', {\bf r}) d\lambda \, ,
\end{equation}
where $\gamma_\lambda( {\bf r},{\bf r}')$ is the one-particle density matrix at coupling $\lambda$.
This follows from expanding the $\delta$ function in one-electron orbitals $\delta({\bf r},{\bf r}')=\sum_{\rm m \in all} \langle {\bf r} | m \rangle
\langle m  | {\bf r}'\rangle$,
and inserting the independent particle approximation
for the polarizability (compare Equs.~\ref{equ:chi} and~\ref{equ:Greens}).
Hence, in Equ. (\ref{equ:acfdt})  the first  line represents the change of the Hartree energy along
the coupling constant integration, whereas the second line accounts for the change of the exchange energy
(for uncorrelated wave functions).

Returning now to  Equ. (\ref{equ:chi}), we note that replacing
$\chi_\lambda$ by $P_\lambda$ in Equ. (\ref{equ:acfdt})
neglects important many body effects related to changes
in the Hartree or exchange potential. In many body perturbation theory, these can be included 
exactly by solving the Bethe-Salpether equation for the polarization propagator.\cite{Hanke_BSE_PRL_1979,Hanke_BSE_1980,Albrecht_BSE_1998,Rohlfing_BSE_1998}
The simplest approximation to the BSE equation is the common RPA approximation
\begin{equation}
\label{equ:dysonP}
\bar \chi_\lambda(\omega) =\bar P_\lambda(\omega)+ \bar P_\lambda(\omega) \lambda v\bar \chi_\lambda(\omega),
\end{equation}
which only includes the Hartree-related ring or ``bubble" diagrams.
In time dependent DFT, the related equation is 
\begin{equation}
\label{equ:dysonPRPA}
\bar \chi_\lambda(\omega) =\bar P_0(\omega)+ \bar P_0(\omega) (\lambda v+f^{\rm xc}(\lambda)) \bar \chi_\lambda(\omega),
\end{equation}
where $f^{\rm xc}(\lambda)$ accounts for all correlation effects,
including the change of the independent particle polarizability
along the coupling path.

If the density
is kept constant   ($n_\lambda=n_0$) and one approximates $P_\lambda(\omega)= P_0(\omega)$
in Equ. (\ref{equ:dysonP}), one obtains for Equ.~(\ref{equ:acfdt}) the 
direct RPA correlation energy:\cite{niquet2003pra}
\begin{equation}
\label{equ:RPA}
 E^{\rm RPA}_c= \frac{1}{2\pi} \int_0^{\infty} d\omega \, \Tr [\ln (1 -\bar P_0(\omega) v) +\bar P_0(\omega) v].
\end{equation}
Here and in the following, we often drop position indices and implicitly assume integration over
the missing spatial coordinates:
\[
 (A B) ({\bf r},{\bf r}'')= \int A({\bf r},{\bf r}') B({\bf r'},{\bf r}'') d^3{\bf r'}.  
\]
\subsection{Singles contribution in G\"orling-Levy perturbation theory}
\label{sec:theory_GL2}
Before considering density changes along the coupling constant integral, we will briefly derive the singles expression
in the  AC-FDT for a fixed charge density in order to 
compare with this equation later.

In standard RPA, one  neglects that  the independent particle polarizability  
changes along the coupling constant integration, 
as the {\em one-particle Green's function
changes as $\lambda$} changes (see Equ. \eqref{equ:dyson0} below). Oddly this
term is not often considered in the AC-FDT framework, although G\"orling and Levy
have already highlighted its relevance (albeit not in a Green's function formalism).
A  partially analogous discussion using the Green's function formalism can be found in e.g. Ref.~\onlinecite{Hellgren_Barth_RPAX_2008,Toulouse_range-separated_2010}.

At small coupling $\lambda$, the change of the one-particle Green's function
is described by the first order term in the Dyson equation
(see e.g. Ref. \onlinecite{Stan_Leeuwen_scGW_2009}):
\begin{equation}
\label{equ:dyson0}
\begin{split}
 G_\lambda(0) & = G_0(0) +   \lambda \int_{-\infty}^{\infty} d \tau\, G_0(\tau) (\Delta \hat V^{\rm KS}+  \hat V^x) G_\lambda(-\tau)  \\
& \approx G_0(0) +  \lambda  \int_{-\infty}^{\infty} d \tau\,  G_0(\tau) (\Delta  \hat  V^{\rm KS}+  \hat V^x) G_0(-\tau) ,
\end{split}
\end{equation}
where $\Delta \hat V^{\rm KS}({\bf r}) \delta({\bf r},{\bf r}')$ is the change of the local Kohn-Sham potential, and 
$\hat V^x({\bf r},{\bf r}')$ is the exact non-local exchange potential evaluated using the DFT orbitals
at coupling constant zero: this term originates from switching on the exact many body potential, which
in lowest order is equivalent to switching on  the non-local exchange. 
If one performs the coupling constant integral  keeping
the density fixed, then the change of the Kohn-Sham potential must be
chosen such that the density remains exactly constant to the original
density at $\lambda=0$.
This requirement can be written as:
\begin{equation}
\label{equ:densGreens}  
n_\lambda({\bf r})-n_0({\bf r})= \lim_{\tau \to 0^-} \left(G_\lambda(\tau, {\bf r},{\bf r})  -  G_0(\tau, {\bf r},{\bf r}) \right)=0,
\end{equation}
as the diagonal of the  Green's function at $\tau \to 0^-$ is just the charge density.
From Eq.~(\ref{equ:dyson0}) one obtains the linearized Sham-Schl\"uter equation for the potential $\Delta V^{\rm KS}$:\cite{ShamSchluter_OEP_1983}
\begin{equation}\label{equ:ShamSchluter0}
 \int_{-\infty}^{\infty} d \tau\,  G_0(\tau) (\Delta V^{\rm KS}+ V^x) G_0(-\tau) =0.
\end{equation}
Since the Kohn-Sham potential is local, one can factor out  $ \bar P_0(\omega=0)= \int d \tau G_0(\tau)G_0(-\tau)$ (see
Equ. \eqref{equ:ft}) and
obtain:
\begin{equation} 
\label{equ:ShamSchluter}
\begin{split}
-\Delta \hat V^{\rm KS}&  = V^{\rm EXX} =   \bar P_0^{-1} (0) \int_{-\infty}^{\infty} d \tau G_0(\tau) V^x G_0(-\tau)\,.
\end{split}
\end{equation}
This is just the local exact-exchange optimized-effective potential (EXX-OEP) $V^{\rm EXX}$.
In summary, if the density is supposed to remain constant
along the coupling constant, the first order change of the local Kohn-Sham
potential is exactly given by the exact-exchange OEP potential. This is not surprising,
since the defining property of the Sham-Schl\"uter equation is 
that the density from the non-local exchange potential and the effective local
potential must equal each other. 

\begin{figure}
    \begin{center}
       \includegraphics[width=8cm,clip=true]{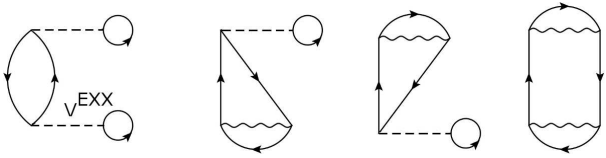}
    \end{center}
   \caption{
Goldstone diagrams corresponding to the singles. The sign is given by the number
of closed Fermionic loops (negative for first and last diagram, positive 
for the two in the middle). The exact exchange potential is density dependent
(bubble) and indicated by a broken line.
}
\label{fig:goldstone}
\end{figure}

One can now calculate the change of the correlation energy by inserting the first order change
of the Green's function into the expression for the correlation energy in the AC-FDT. 
Approximating $\chi_\lambda \to P_\lambda=G_\lambda G_\lambda$, inserting (\ref{equ:chi}) 
and (\ref{equ:dyson0}) into the second line of Equ.~(\ref{equ:acfdt}), and rewriting the $\delta$ function as
a sum over states, or alternatively, starting from Equ. (\ref{equ:acfdt_ex})
and identifying $\gamma_\lambda= G_\lambda(0^-)$ yields 
a change of the correlation energy of (yet omitting the integration over $\lambda$):
\begin{equation}\label{equ:aux1}
\begin{split}
  -\frac{\lambda}{2} \int_{-\infty}^{\infty} & \left( \Tr [G_0(\tau) (\hat
V^x- \hat V^{\rm EXX}) G_0(-\tau) G_0(0^-) \hat v \right.\\
                       +&\left. \Tr[G_0(0^-) G_0(\tau) (\hat V^x- \hat V^{\rm
EXX}) G_0(-\tau)  \hat v ] \right)\,d \tau.
\end{split}
\end{equation}
Because of the trace, the second term yields just the complex conjugate of the first term. Furthermore, the integral over
the positive half-plane of $\tau$ gives the same value as that over the negative half-plane ($\tau<0$). Finally,  
the term $- G_0(0^-) \hat v$ corresponds to the non-local exchange potential $\hat V^x$.
Inserting the defining equation for the one-particle Green's function (\ref{equ:Greens})
and performing the integration over $\tau$ and $\lambda$
yields what is commonly referred to as the singles contribution:
\begin{equation}
 E_c^{\rm SE}=-\sum_{\substack{a\in{\rm virt},\\i\in{\rm occ}}}
 \frac{ | \langle i | \hat V^x- \hat V^{\rm EXX}  |  a \rangle|^2} {\epsilon_a - \epsilon_i}\,.
\end{equation}
Note that one needs to add Equ. (\ref{equ:ShamSchluter0}) to Eq. \eqref{equ:aux1} 
multiplied by $\hat
V^{\rm EXX}({\bf r})$ and
integrated over $\bf r$  to derive this convenient equation.
The corresponding (time-ordered) Goldstone diagrams are also shown schematically in Fig. 
\ref{fig:goldstone}.
A few comments are in place here. The term has been first derived by
G\"orling and Levy.\cite{Goerling_Levy_GL2_1994} In their derivation it is, however,
not obvious that this term describes the changes of the correlation energy from  the  one-particle
density matrix (which equals the one particle Green's function at $\tau=0^-$) along the coupling constant 
integral. In fact, most standard
AC-FDT calculations neglect the term. Only the RPAX (RPA including eXchange) following G\"orling  and coworkers
accounts for this term (often referred to as EXX-RPA).\cite{Hesselmann_RPA-EXX_2011} 
In these formulations, however, the change of the Green's function is accounted for by
recasting it (as well as the particle-hole ladder diagrams) into an effective exchange 
kernel $f^{xc}(\lambda) \approx \lambda f^x$ for the polarizability (compare Equ. \ref{equ:dysonPRPA}).
\cite{Hellgren_Barth_RPAX_2008,Hesselmann_RPA-EXX_2011}

We conclude, {\em  the singles account for  the change of the density matrix along
the coupling constant integral}. For a constant density,  obviously only changes
of the exchange and kinetic energy can be included. In the next step, we will also
allow for changes of the charge density along the coupling constant
integral.

\subsection{Singles contribution with density changes}
\label{sec:theory_singles}

We now derive the singles contribution to the correlation energy for the case
when the density does not stay constant during the coupling constant integration.
Changes of the charge density are most easily accounted for by linearly switching
off the Kohn-Sham potential (see Equ. \eqref{equ:linear}) and linearly switching on the
exact many body interaction.
In principle, this makes the derivation even less involved, since the determination of the local
exact exchange potential is no longer required. We first again derive the
expression  in lowest order, where  the Green's function is now
given by 
\begin{equation}
\label{equ:dyson2}
 G_\lambda(0) \approx G_0(0) + \lambda  \int_{-\infty}^\infty d \tau'\,  G_0(\tau') (\hat V^{\rm HF} - \hat V_0^{\rm KS}) G_0(-\tau').
\end{equation}
In the lowest order, the change of the potential is now the difference between 
the Hartree-Fock potential $\hat V^{\rm HF}= \hat V^{\rm H} +  \hat V^{x}$,
the sum of the Hartree and exchange potential,  
and the original Kohn-Sham potential, which is adiabatically switched off.
As before, the change of the density matrix is given by the change of the
Green's function $G(\tau \to 0^-)$ (compare Equ.~\eqref{equ:densGreens}).
In this case, the first and second line of Equ.~(\ref{equ:acfdt}) can be combined to yield:
\begin{equation}
 \frac{\lambda}{2} \int_{-\infty}^{\infty} d \tau\, \Tr [G_0(\tau) (\hat V^{\rm HF}- \hat V_0^{\rm KS}) G_0(-\tau) \hat V^{\rm HF}]+ c.c.  
\end{equation}
The first line of Equ.~(\ref{equ:acfdt}) yields the Hartree-potential times
the change of the density, whereas the second line yields the  exchange potential times
the change in the density matrix,  in sum  the change of the density matrix times $\hat V^{\rm HF}$.

After performing the coupling constant integral this yields
\begin{equation}
  E_c= -\sum_{\substack{a\in{\rm virt},\\i\in{\rm occ}}}
 \frac{ \langle i | \hat V^{\rm HF}- \hat V_0^{\rm KS}  |  a \rangle \langle a | \hat V^{\rm HF} | i\rangle } {\epsilon_a - \epsilon_i}.
\end{equation}
This term needs to be combined with the term $\Delta E_c$ given in Equ.~(\ref{equ:ccintegral_rho}), which
can be calculated by inserting Equ.~(\ref{equ:dyson0}) 
and performing the $\tau$ and $\lambda$ integration.
Both contributions combined yield a simple term 
\begin{equation}
\label{equ:rSE_ren}
 E^{\rm SE}_c +  \Delta E_c = - \sum_{\substack{a\in{\rm virt},\\i\in{\rm occ}}}
 \frac{ |\langle i | \hat V^{\rm HF}- \hat V_0^{\rm KS}  |  a \rangle|^2 } {\epsilon_a - \epsilon_i},
\end{equation}
which exactly corresponds to the singles suggested by Ren et al.\cite{ren2011,Ren_Singles_2012}
Here, we have performed the derivation concisely within
the AC-FDT framework instead of Rayleigh-Schr\"odinger perturbation theory,
and, as it must be, both are exactly equivalent. As in the previous paragraph, the singles
account for the change of the mean field exchange energy. However, now they also {\em include the change of the 
mean field Hartree energy} along the coupling constant integration. Here and in the following,  
we define the {\em mean field} as the contributions that stem from the one-particle Green's function
and the related density matrix. 

In the renormalized singles of Ren and coworkers\cite{ren_renormalized_2013} also higher order
contributions are accounted for.
However, it is not entirely straightforward to generalize our results to include higher order terms in $\lambda$,
and to continue, we make one crucial approximation. Let us introduce this approximation for
the density term, which can be written as (compare Equ.~\eqref{equ:acfdt})
\begin{equation}
\begin{split}
 \frac{1}{2} &\left( [n_\lambda \hat v n_\lambda ] - [n_0 \hat v n_0 ] \right)  =  \\
\frac{1}{2}  & [(n_\lambda- n_0) \hat v (n_\lambda+n_0) ] \approx [(n_\lambda - n_0)\, \hat v \, n_0 ]. 
\end{split}
\end{equation}
In the second line, we assume that the differences between $n_\lambda$ and $n_0$ are small
so that we can approximate the sum by $2 n_0$. Analogous manipulation is possible
for the term involving the polarizability, if the full polarizability is approximated by the independent
particle approximation Equ. (\ref{equ:chi}) ($\chi_\lambda \to P_\lambda=G_\lambda G_\lambda$). After collecting all terms, adding Equ. (\ref{equ:ccintegral_rho}), and noting
that $n({\bf r})= G(0^-,{\bf r},{\bf r})$, one obtains the approximate
renormalized singles correlation energy 
\begin{equation}
 \approx \int_0^1  \Tr [ (G_\lambda(0^-) - G_0(0^-) ) (\hat V^{\rm HF}- \hat V_0^{\rm KS})] d\lambda \,.
\end{equation}
It is fairly straightforward to backtrack
this into a simple total energy relation (essentially reversing the steps introduced
in Sec.~\ref{sec:theory_ACFDT}).  We first note that the difference between
the Hartree-Fock potential and the Kohn-Sham potential equals the difference
in the corresponding one-electron Hamiltonians $\hat h^{\rm HF}- \hat h^{\rm
KS}$. Next, the constant term $G_0(0^-)$
is integrated over  so that we obtain
\begin{equation}
\label{equ:ccG}
\begin{split}
  \int_0^1  & \Tr [ G_\lambda(0^-) (\hat h^{\rm HF}- \hat h^{\rm KS})] d\lambda \\
  & -\Tr [G_0(0^-) (\hat h^{\rm HF}- \hat h^{\rm KS})].
\end{split}
\end{equation}
As the one-electron Green's function  $G_\lambda(0^-)$ is the exact Green's function
of the one-electron Hamilton operator  $\hat h_\lambda = \hat h^{\rm KS} +\lambda (\hat h^{\rm HF}-\hat h^{\rm KS})$,
one can rewrite the first line as (Hellman-Feynman theorem):
\[
\int_0^1  \Tr \left[ G_\lambda(0^-) \frac{d \hat h_\lambda}{d\lambda} \right] d\lambda 
= \Tr\left[G_1(0^-)\hat h_1 - G_0(0^-)\hat h_0\right].
\]
Combining this with the second line in Equ.~\eqref{equ:ccG} ($\hat h_0=\hat h^{\rm KS}$, $\hat h_1=\hat h^{\rm HF}$)
yields
\begin{equation}
\label{equ:rSE}
 E^{\rm rSE}_c = \Tr[ \gamma_{\rm HF} \hat h_{\rm HF} - \gamma_{\rm DFT}\hat h_{\rm HF}]. 
\end{equation}
Here $\gamma_{\rm HF}=G_{\rm HF}(0^-)$ is the Hartree-Fock density matrix, determined for the
Hartree-Fock Hamiltonian $\hat h_{\rm HF}$, where the Hartree-Fock potential is set up with DFT-orbitals. This is exactly
the ``single-shot'' Hartree-Fock energy: it can be calculated by diagonalizing the
HF Hamiltonian (set up with DFT-orbitals), summing the eigenvalues of the occupied states
and subtracting the original diagonal part of the HF Hamiltonian for
the occupied manifold evaluated using the original DFT orbitals ($i^{\rm DFT}$):
\begin{equation}
\label{equ:rSE_sum}
E^{\rm rSE}_c = \sum_i \epsilon_i^{\rm HF} - \langle i^{\rm DFT} | \hat h_{\rm HF} | i^{\rm DFT} \rangle.
\end{equation}
This prescription is one central result of the present paper. 

In all practical calculations, we found the value of the single shot HF energy 
to be exceedingly close to the renormalized singles introduced by Ren and coworkers.\cite{ren_renormalized_2013} 
For diamond the difference is on the order of 0.3~meV (with the singles being of the order
of 0.3~eV). Even for small band gap systems, such as metallic Pd or Al, differences hardly ever
exceed 1~\% and are entirely irrelevant when evaluating relative energies or
lattice constants. 
The relation of our equation to the singles of  Ren~{\em et al.} is fairly straightforward to see.
Ren~{\em et al.} essentially renormalizes
the propagator $G$ in the occupied-occupied block as well as the unoccupied-unoccupied block,
by diagonalization of these sub-blocks using the HF Hamiltonian $\hat h_{\rm HF}$.
This step is crucial, since the one-electron eigenvalues are renormalized
to the HF-eigenvalues; if LDA eigenvalues were  used in the evaluation of
the singles in Equ. (\ref{equ:rSE_ren}), the response of the system
to the change of the potential from Kohn-Sham  to HF would be strongly overestimated.
Ren then calculates the change of the mean field energy in second order
with the DFT eigenvalues in Equ. (\ref{equ:rSE_ren}) replaced by the
renormalized HF eigenvalues. 
As opposed to this prescription, we
also ``renormalize'' the propagator in the occupied-unoccupied block. 
With the present derivation at hand, there is no obvious reason why not to
chose the simpler prescription of the present work. 
Since results obtained using the renormalized singles (rSE) are in practice  
indistinguishable from results obtained using the single-shot HF
energy change, all calculations here use the single-shot HF energy change,
but are nevertheless labeled as ``rSE".

\subsection{Singles contribution beyond the Hartree-Fock description}
\label{sec:singles_beyond_HF}

An obvious extension to the renormalized singles approach is to use the full RPA
density matrix instead of the HF density matrix to estimate the change
of the mean field energy:
\begin{equation}
\label{equ:GWSE}
 E^{\rm GWSE}_c = \Tr[ \gamma_{\rm RPA}\hat h_{\rm HF} - \gamma_{\rm DFT}\hat h_{\rm HF}], 
\end{equation}
where the $\gamma_{\rm RPA}$ is the RPA density matrix.
In the singles derived in the previous section, it was assumed
that the density matrix of the interacting system is well approximated
by the Hartree-Fock case, which seems a crude approximation.
Given the excellent performance of the RPA for total energies and
band gaps, the RPA density matrix, however, should approximate
the density matrix of the real interacting system very accurately.
At this point, we have, however, no entirely concise derivation for the term,
although the physical interpretation is clear:
it should account approximately for the change of the mean-field kinetic, Hartree and exchange 
energy  along the coupling constant integral.

To  evaluate the RPA density matrix, we calculate 
the RPA Green's function
\begin{equation}
\label{equ:dyson}
 G_{\rm RPA}(i \omega) = G_0(i \omega) +  G_{\rm RPA}(i\omega)  (\Sigma(i\omega)-\hat V_{xc}^{\rm KS}) G_0(i\omega) \, .
\end{equation}
Here $\Sigma$ is the self-energy in the $GW$ or random phase approximation
(the two approximations are synonymous)
\begin{equation}
 \label{equ:sigma}
  \Sigma(\tau,{\bf r}, {\bf r}')= -G_0(\tau,{\bf r}, {\bf r}')W(\tau, {\bf r}', {\bf r}),
\end{equation}
and $G_0$ and $W$ are the Kohn-Sham Green's functions (\ref{equ:Greens}) and the screened potential evaluated using
Kohn-Sham polarizabilities.\cite{Hedin_GW_1965}  $\hat V_{xc}^{\rm KS}$ is the exchange-correlation
contribution to the Kohn-Sham potential.

To evaluate the RPA density matrix  numerically, we determine the interacting Green's function 
at full coupling $G_{\rm RPA}(i \omega)$ using Equ.~(\ref{equ:dyson}),
transform it to the imaginary time at a small negative infinitesimal $\tau \to 0^-$
to obtain the one-particle density matrix for the RPA, $\gamma_{\rm RPA}({\bf r}, {\bf r}')= G_{\rm RPA}(0^-, {\bf r}, {\bf r}')$,
and finally diagonalize the density matrix to obtain the natural orbitals
\[
\gamma_{\rm RPA} = \sum_m | m \rangle f_m  \langle m |.
\]
In the present work, we use one more crucial approximation:
instead of using the occupancies of the actual density matrix,
we keep the original occupancies evaluated on the level of density
functional theory ($1$ and $0$ for insulators). This has
two reasons: first the density matrix evaluated from the
Green's function (\ref{equ:dyson}) is not particle conserving,
{\em i.e.} the number of electrons deviates from the original number.\cite{Stan_Leeuwen_scGW_2009}
Only if the Green's function in (\ref{equ:dyson}) and (\ref{equ:sigma}) is iterated
to self-consistency, the particle number is conserved. 
To test our present code, we have also iterated
the Green's function in both equations to selfconsistency, and in that case, the electron
number is indeed exactly conserved. However, such
calculations are fairly expensive and difficult to apply routinely. 
They would also most likely require us to  combine it with a different treatment
of the fluctuation terms beyond the standard RPA treatment as used here.\cite{Stan_Leeuwen_scGW_2009}

The second reason is based on the definition of singles in
quantum chemistry. 
A density matrix with occupancies $0$ and $1$ corresponds to a single
Slater determinant. We, therefore, approximate the density matrix
by the ``best'' single Slater determinant approximating the correlated 
RPA density matrix. This is 
consistent with  Brueckner coupled cluster orbitals,\cite{Stolarczak_Brueckner_CC_1984}
which are obtained by performing a rotation between the occupied and unoccupied manifold
to determine an optimal reference single Slater determinant. The rotation is
chosen to remove all ``singles'' contributions from the correlation energy. 
In quantum chemistry, the fluctuations
that cause the fractional occupancies in the density matrix are, in fact,
not included in the singles. For instance,
in the coupled cluster theory,  fluctuations are accounted for by double excitation operators
(doubles). Per definition, singles are diagrams ending
in a single excited determinant with one hole in an orbital $i$ and an
additional electron in an orbital $a$ (compare Fig.~\ref{fig:goldstone}).  
We essentially follow the quantum chemistry  convention
in partitioning the correlation energy into a ``singles" term and
fluctuation terms described entirely  by Equ.~(\ref{equ:RPA}).
We finally note that the RPA orbitals constructed in this manner are similar 
but not identical with the  Brueckner RPA direct ring coupled-cluster orbitals suggested 
by Moussa.\cite{Moussa_cubic_RPA_2014}

\section{Computational setup}

\begin{table}
\caption{
PAW Potentials and plane wave cutoffs for the orbital basis used for the individual calculations
(as distributed with {\tt vasp.5.4.1}).
The shorthand, {\tt $\_$sv} indicates that the upper most core shell was included
as ``valence'' orbitals (see explanation in Sec. \ref{sec:solids}).
For the norm-conserving case, the enumerated potentials were replaced by the same potential
with an appendix {\tt $\_$nc}, and for N,  O, and F the small core potentials  N$\_$h$\_$GW, O$\_$h$\_$GW,  and 
F$\_$h$\_$GW were used. These potentials are almost norm-conserving.
For the norm-conserving potentials always the default cutoff was used.
}
\label{tab:setup}
\begin{ruledtabular}
\begin{tabular}{llc}
     &   applied potentials &   cutoff (eV)    \\
\hline     
Ne & Ne$\_$GW & 1000  \\
Ar & Ar$\_$GW &  600  \\
Kr &  Kr$\_$GW &   500 \\
ice & O$\_$GW H$\_$GW & 600 \\
benzene & C$\_$GW H$\_$GW & 600\\
water on NaCl &  Na$\_$sv$\_$GW Cl$\_$GW  O$\_$GW  H$\_$GW & 400 \\                                                    
Na   &   Na$\_$sv$\_$GW            &  350  \\ 
Al   &   Al$\_$sv$\_$GW            &  534  \\ 
Rh   &   Rh$\_$sv$\_$GW            &  351  \\ 
Pd   &   Pd$\_$sv$\_$GW            &  356  \\ 
Cu   &   Cu$\_$sv$\_$GW            &  509  \\ 
Ag   &   Ag$\_$sv$\_$GW            &  460  \\ 
C    &   C$\_$GW$\_$new            &  414  \\ 
Si   &   Si$\_$GW               &  245  \\
Ge   &   Ge$\_$sv$\_$GW            &  533  \\
LiF  &   Li$\_$sv$\_$GW F$\_$GW       &  487  \\
LiCl &   Li$\_$sv$\_$GW Cl$\_$GW      &  433  \\
NaF  &   Na$\_$sv$\_$GW F$\_$GW       &  487  \\
NaCl &   Na$\_$sv$\_$GW Cl$\_$GW      &  341  \\
MgO  &   Mg$\_$sv$\_$GW O$\_$GW       &  434  \\
SiC  &   Si$\_$GW C$\_$GW$\_$new      &  414  \\
AlN  &   Al$\_$sv$\_$GW N$\_$GW$\_$new   &  547  \\
AlP  &   Al$\_$sv$\_$GW P$\_$GW       &  534  \\
AlAs &   Al$\_$sv$\_$GW As$\_$sv$\_$GW   &  539  \\
GaN  &   Ga$\_$sv$\_$GW N$\_$GW$\_$new   &  420  \\ 
GaP  &   Ga$\_$sv$\_$GW P$\_$GW       &  404  \\
GaAs &   Ga$\_$sv$\_$GW As$\_$sv$\_$GW   &  539  \\
InP  &   In$\_$sv$\_$GW P$\_$GW       &  476  \\ 
InAs &   In$\_$sv$\_$GW As$\_$sv$\_$GW   &  539  \\ 
InSb &   In$\_$sv$\_$GW Sb$\_$sv$\_$GW   &  484  \\  
\end{tabular}
\end{ruledtabular}
\end{table}

 
The present calculations were performed using the VASP code.\cite{Kresse_VASP_1996}
We used the new implementation of the RPA routines as documented
in Refs. \onlinecite{Kaltak_LowScaling_2014,Kaltak_SiLowScaling_2014}.
The code has been extended to allow for (self-consistent) calculations
of the one particle Green's function in the random phase approximation
{\em i.e. } solving  Equ.~(\ref{equ:dyson}) and~(\ref{equ:sigma}). 
Here only single shot calculations are performed, by calculating
the self-energy once at full coupling  and determining  
the RPA Green's function and RPA density matrix (Equ.~\eqref{equ:dyson}).
The singles contributions are evaluated either using the single
shot HF density matrix (Equ.~\eqref{equ:rSE}) or
the single shot RPA density matrix (Equ.~\eqref{equ:GWSE}).
We term the two results either rSE (renormalized singles) or
GWSE, respectively. For the fluctuation term we use the standard random phase approximation  as defined in Equ.~(\ref{equ:RPA}). 
Furthermore, we often show the results of the random phase approximation (evaluated using
DFT orbitals) combined with HF exchange energies. In this case, the exact exchange energy evaluated
using DFT orbitals is replaced by the selfconsistent HF energy.
The DFT orbitals are always determined 
using the PBE functional, where  PBE stands for the usual Perdew-Burke-Ernzerhof
functional.\cite{perdew1996} 
The potentials are generally equivalent to the GW potentials
distributed with {\tt vasp.5.4.1}. If not otherwise stated
the default plane wave cutoffs were used for the calculations. 
For the response function, the cutoff was set to 
2/3 of the cutoff for the orbitals. If the energy-volume curves were
non smooth using $4\times 4 \times 4$ k-points, the plane wave
energy cutoffs were increased by 30~\%, roughly doubling the
number of included plane waves. The applied computational setup
and potentials are summarized in Tab.~\ref{tab:setup}.

To determine the equilibrium volumes, the volumes were typically varied in steps
of $5 \%$ by $\pm 15 \%$ around the experimental equilibrium volume, and the
Murnaghan equation of state was fitted to the data points.\cite{murnaghan1944}

\section{Results}
\label{sec:results}

\subsection{Rare gas solids}
\label{sec:raregas}

\begin{table}
\caption{
Lattice constants (in \AA)  as well as cohesive energies (in meV)
of rare gas solids calculated using the RPA and different variants for the kinetic,
exchange and Hartree energy contributions.
The experimental data have been corrected for zero point vibration energies
as determined by accurate quantum chemical methods.\cite{Paulus_CC_2000}
}
\label{tab:raregas}
\label{}
\begin{ruledtabular}
\begin{tabular}{lrrrrr}
volumes& EXX+RPA & +rSE      & +GWSE   & HF+RPA     & EXP    \\
Ne    &  4.35    &  4.32     & 4.38    &  4.28      & 4.30  \\
Ar    &  5.33    &  5.24     & 5.25    &  5.20      & 5.25  \\
Kr    &  5.68    &  5.61     & 5.61    &  5.56      & 5.63  \\
\hline
energies& EXX+RPA&  +rSE & +GWSE       & HF+RPA     & EXP \\
Ne    &  19      &   35     &   30     &   36       &  26.2  \\ 
Ar    &  69      &   88     &   87     &   93       &  87.9  \\
Kr    &  97      &  119     &  119     &  126       & 121.8  \\
\end{tabular}
\end{ruledtabular}
\end{table}

Rare gas solids constitute a prototypical test case for
van der Waals bonded solid state systems. Although, the standard RPA
performs reasonably well for rare gas solids,\cite{harl2008} one observes
that the binding energy is quite significantly  underestimated,
in particular for He. Inclusion of singles has remedied
this issue for the rare gas dimers, and one would expect that
this also applies to solids.\cite{ren2011,ren_renormalized_2013}

We first note that quantum chemical coupled cluster calculations
at the level of CCSD(T) (coupled cluster with singles and doubles
and perturbational triples) using the method of increments
and including up to four body interactions yield essentially
exact results within a few $\mu$Hartree.\cite{Paulus_CC_2000,Schwerdtfeger_CC_2010} 
To compare with experiment we have corrected the experimental data
for zero point vibration energies in both the cohesive energy,
as well as in the lattice constants.\cite{Paulus_CC_2000}

The present calculations have been performed using relatively
high plane wave energy cutoffs of 1000~eV, 600~eV and 500~eV
for Ne, Ar and Kr in order to obtain smooth energy-volume curves.
$4\times 4 \times 4$ k-points were used, although already
$3\times 3 \times 3$ k-points yield practically identical results.
The PAW potentials are approximately norm-conserving to 
avoid errors in the vdW contributions from excitations into high lying unoccupied
orbitals.\cite{klimes_NCPAW_2014} As already mentioned, standard RPA combined with exact exchange yields at best
modest agreement with experiment (see Tab. \ref{tab:raregas}). Specifically, the equilibrium 
volumes are overestimated and the binding energies are, as
already noted above, about 20~\% too small, with the errors
being particularly large for Ne. 
Inclusion of the singles contributions from HF (rSE)
and using the RPA density matrix (GWSE) yields
a clearly improved agreement with experiment in particular for
Ar and Kr. Ne is less satisfactory.
For Ne, the binding energy significantly overshoots if the 
singles are evaluated on the Hartree-Fock level (rSE). This improves,
if the RPA density matrix is used, however, including GW-singles worsens
the volume even compared to RPA. We believe that
this is a result of the PBE approximation being
particularly inaccurate for large band gap system such as Ne.
This is for instance exemplified by the fact that the binding
energy almost doubles, when the exact exchange 
evaluated using DFT orbitals is replaced by the exact
exchange evaluated using Hartree-Fock orbitals (HF+RPA).
For Ar and Kr the changes are typically only 25~\%. Also other metrics
indicate that the error in the orbitals is significant
using the PBE functional for this case. Hence, single shot
corrections, be it RPA+rSE or RPA+GWSE, yield less reliable results then for
Ar and Kr. For Ar and Kr, the performance of RPA+rSE and RPA+GWSE
is remarkably good, approaching that of state of
the art quantum chemistry methods. Furthermore, the differences between
rSE and GWSE are small (except for Ne), which is
related to the fact that all these systems screen relatively 
weakly. Therefore, the RPA density matrix is very close
to the HF density matrix (see below).

\subsection{Ice}
\label{sec:ice}

Ice is another system, where singles are expected to have
a significant impact.  As for rare gas solids, the
PBE charge density is too spread out and replacing the exact
exchange evaluated using PBE orbitals by the exact Hartree-Fock
exchange increases the binding energy by 100 meV.\cite{Macher}
The predicted volumes using EXX+RPA and HF+RPA almost exactly bracket
the experimental values.

Here we only concentrate on one ice phase, the lowest energy cubic phase
of ice, I$_c$(a), with a ferroelectric order. In our previous study,
we found this phase to be practically iso-energetic with the ferroelectrically
ordered ice XI in the  Cmc2$_1$  space group. Ice XI is a
proton ordered variant of the common proton disordered phase of hexagonal ice.
In the present study, we used  PBE relaxed structures and identical potentials as in our previous
study.\cite{Macher} However, the cutoff was increased from 450 eV
to 600 eV. This generally results in smoother energy-volume curves and
changes the calculated volumes by about   0.5~\%: 
because of the increased cutoff and reduced noise in the calculated data, 
the present calculations are slightly more accurate.

As already observed in our recent work, the EXX+RPA underbinds, whereas
combining the RPA with Hartree-Fock energies overestimates the binding
energies (see Tab.~\ref{tab:ice}). Including the singles in the HF approximation improves
the description already significantly, although the results
are still too close to the HF case. In this case, the singles
evaluated using the RPA-density matrix (GWSE) yield results very close
to experiment and practically identical to the diffusion
Monte Carlo data, which gives an equilibrium volume per molecule of
31.7 \AA$^3$ and a binding energy of 605 meV
for hexagonal ice.\cite{Santra_Ice_PRL_2011} Since
hexagonal ice has a 0.2 \AA$^3$ smaller volume and 
a 5 meV reduced  binding energy due to disorder
(compare Table III in Ref.~\onlinecite{Macher}), the results
are virtually on top of the DMC data at a tiny fraction
of the compute cost.

\begin{table}
\caption{
Equilibrium volume per molecule  (in \AA$^3$)  as well
as cohesive  energy (in meV) with respect to an isolated H$_2$O molecule
for cubic ferroelectrically ordered ice I$_c$(a).
}
\label{tab:ice}
\begin{ruledtabular}
\begin{tabular}{lrrrrr}
          & EXX+RPA  & +rSE       & +GWSE   & HF+RPA    &   EXP     \\
volume    &  32.94   &   31.76    & 32.03   &  31.38    &     32.105$^a$ \\
energy    &  536     &   630      & 620     &  670      &    610  \\
\end{tabular}
\end{ruledtabular}
\begin{flushleft}
$^a$ Ref. \onlinecite{Vega_ice_TIP_2005}
\end{flushleft}
\end{table}
\subsection{Benzene}
\label{sec:benzene}

Benzene is the simplest aromatic molecule and therefore it represents 
an interesting reference point for the study of more complex molecular solids.
The delocalized character of $\pi$ bonds in benzene makes an accurate calculation
of interactions or cohesive energies difficult, both, in the gas phase, as well as  in the condensed phases.
Simple methods, such as MP2, lead to overestimated binding energies and 
more involved schemes need to be used to obtain accurate results.
For example, for crystalline benzene Wen and Beran computed a post-MP2 correction of 10.4~kJ/mol (108~meV),\cite{wen2011jctc}
reducing the MP2 cohesive energy considerably.
Li~{\it et al.} estimated the RPA cohesive energy to be 47~kJ/mol or 487~meV,\cite{li2010rpa}
significantly below the recently revised experimental estimate of $-55.3\pm2.2$ kJ/mol (about 573~meV).\cite{yang2014}

We have used the geometry of the crystal and monomer optimised with the optB88-vdW functional
and performed both the extrapolation to infinite k-point mesh and infinite basis sets.
To obtain the reference energy of the monomer, we also performed calculations at different
volumes and extrapolated to the infinite cell volume.
The RPA and EXX calculations used up-to 3$\times$3$\times$3 and 6$\times$4$\times$6 k-point 
meshes, respectively.

The data are summarised in Table~\ref{tab_bz}. 
Our EXX+RPA value is similar to the previous calculation of Li~{\it et al.} and 
underestimates the reference by about 110~meV.
The effect of the single excitations is rather small, the binding energy increases by $40$~meV
for the HF based singles and by about $55$~meV when we use the GW singles.
This means that the final error is halved compared to the original RPA calculation.
The hybrid scheme, where self-consistent Hartree-Fock is used for the mean field
part, gives results in a better agreement with the reference data, underestimating it by
about 43~meV.
We note that such a performance is in agreement with the results of Ren~{\it
et al.},\cite{ren2012}
who found that RPA underestimates interaction energies for benzene dimers both
in the stacked and in the T-shaped geometries.
Moreover, including single excitations did not improve the results substantially.
This clearly shows that singles have limitations. We believe that the main error
in this case stems from an inaccurate description of the delocalized  $\pi$ electrons in PBE.

\begin{table}
\caption{
Cohesive energies of benzene crystal obtained with different methods and compared to the
reference value derived from experimental data. }
\label{tab_bz}
\begin{ruledtabular}
\begin{tabular}{lc}
Method & $E_{\rm coh}$ (meV) \\
\hline
EXX+RPA   & $-463$ \\
+rSE      & $-503$\\
+GWSE     & $-518$ \\
HF+RPA    & $-520$ \\
Reference & $-573\pm23$ \\
\end{tabular}
\end{ruledtabular}
\end{table}

\subsection{Water on NaCl}
\label{sec:water}

To explore the accuracy of the RPA and the singles corrections
for adsorption on surfaces,  we have calculated the adsorption energy of a water molecule on 
the NaCl surface.
This is a prototypical system and one for which an estimate
of the adsorption energy of water has been obtained.\cite{li2008}
However, this reference data was calculated using an embedded cluster approach
to obtain the correlation energy. This makes a direct comparison to our
calculations, that are necessarily a finite coverage, difficult.
Therefore, we opted not to use this reference data and instead obtained an estimate from MP2 calculations
for a small system.
To reduce the computational cost of MP2, we, furthermore, restrict the study to a small p(1$\times$1) supercell
with a single water molecule and two layers of NaCl and use that consistently for both MP2 and RPA. 
This corresponds to a high coverage, with one molecule per one surface sodium atom.
Moreover, the reference surface and isolated molecule have the same geometry as in the 
adsorbed structure, and the same simulation cell is used for all cases 
to allow for efficient error compensation. 

The interaction energy depends only weakly on the cutoffs chosen for orbital and 
auxiliary plane-wave basis sets. 
For MP2 we set the cutoff for the orbital basis  and auxiliary polarizability basis to 350~eV 
and  450~eV, respectively.
The interaction energy depends also only weakly on the k-point sampling, we have used
up to 3$\times$3$\times$1 k-points (the cutoff for the auxiliary basis in this case
was 250~eV). 
After accounting for basis set convergence and k-point convergence, we obtained an 
estimate of the molecule-surface interaction energy of $E_{\rm int}=-420$~meV.
Corrections beyond MP2 are expected to be small. In the work of Li~{\it et al.}\cite{li2008},
 CCSD(T) calculations lead to a 10~meV stronger binding energy than MP2.
Further corrections will arise from the use of hard PAW potentials, for example, 
the RPA binding increases by 15~meV when small core potentials are used, but since
we compare MP2 and RPA with similar setups this is irrelevant in the present case.

Our results are summarized in Table~\ref{tab_salt}. As a reference, we use our MP2 value
corrected with the post-MP2 correction of Li~{\it et al.}, which yields,  in total, 
a binding energy of $-430$~meV.
As expected, RPA underestimates this value, by about $50$~meV.
Adding the singles correction (rSE) leads, in this case, to a perfect agreement with the estimated reference
data. 
This is also in agreement with the calculations of Ren~{\it et al.} performed on the S22 
test set.\cite{ren2012}
They found a very good agreement between the reference data and RPA+rSE
for those dimers in S22 that are bound by hydrogen bonds.
The hybrid HF+RPA overestimates the reference interaction energy, which is also in agreement
with the findings of Ren~{\it et al.}
In this case, the singles in the GW approximation (GWSE) do not quite work
as well, but the agreement with the reference data is still reasonable.

\begin{table}
\caption{
Interaction energies of water molecule with NaCl(100) at a high coverage as calculated
by different methods.
}
\label{tab_salt}
\begin{ruledtabular}
\begin{tabular}{lc}
Method & $E_{\rm int}$ (meV) \\
\hline
MP2     & $-420$ \\
MP2+estimated CC correction & $-430$ \\
EXX+RPA & $-383$ \\
+rSE    & $-430$ \\
+GWSE   & $-410$ \\
HF+RPA  & $-450$ \\
\end{tabular}
\end{ruledtabular}
\end{table}

\subsection{Covalent and metallic solids}
\label{sec:solids}

\begin{table*}
\caption{
Lattice constants (in \AA) of selected semiconductors, insulators and metals.
The zero-point vibration corrected experimental lattice constants
are taken from Ref.~\onlinecite{Schimka_HSEsol_2011}.
The table also summarizes the mean relative errors (MRE) and the mean absolute relative errors (MARE)
in percent.
}
\label{tab:lattice}
\begin{ruledtabular}
\begin{tabular}{lcccccc}
     &     RPA    & RPA     &    +rSE  &    +GWSE  &  exp    & \\
     &   NC-PAW   & std-PAW & std-PAW  &   std-PAW &         & \\
\hline                                                        
Na   &      4.140 &   4.195 &   4.188  &    4.200  &   4.214 & \\ 
Al   &      4.022 &   4.034 &   4.016  &    4.020  &   4.018 & \\ 
Rh   &      3.808 &   3.807 &   3.818  &    3.807  &   3.794 & \\ 
Pd   &      3.895 &   3.893 &   3.929  &    3.893  &   3.876 & \\ 
Cu   &      3.601 &   3.600 &   3.659  &    3.605  &   3.595 & \\ 
Ag   &      4.051 &   4.074 &   4.086  &    4.073  &   4.062 &\\ 
C    &      3.562 &   3.571 &   3.577  &    3.575  &   3.553 &\\ 
Si   &      5.428 &   5.438 &   5.445  &    5.445  &   5.421 &\\
Ge   &      5.623 &   5.634 &   5.635  &    5.632  &   5.644 &\\
LiF  &      3.994 &   3.993 &   3.998  &    3.993  &   3.972 &\\
LiCl &      5.076 &   5.079 &   5.071  &    5.073  &   5.070 &\\
NaF  &      4.551 &   4.607 &   4.615  &    4.612  &   4.581 &\\
NaCl &      5.547 &   5.588 &   5.576  &    5.580  &   5.569 &\\ 
MgO  &      4.200 &   4.217 &   4.230  &    4.226  &   4.189 &\\
SiC  &      4.353 &   4.367 &   4.374  &    4.373  &   4.346 &\\
AlN  &      4.366 &   4.382 &   4.388  &    4.388  &   4.368 &\\
AlP  &      5.460 &   5.470 &   5.474  &    5.473  &   5.451 &\\
AlAs &      5.646 &   5.666 &   5.668  &    5.663  &   5.649 &\\
GaN  &      4.493 &   4.508 &   4.510  &    4.508  &   4.509 &\\ 
GaP  &      5.442 &   5.446 &   5.459  &    5.462  &   5.439 &\\
GaAs &      5.620 &   5.643 &   5.641  &    5.634  &   5.640 &\\
InP  &      5.869 &   5.871 &   5.888  &    5.887  &   5.858 &\\ 
InAs &      6.036 &   6.062 &   6.088  &    6.089  &   6.049 &\\
InSb &      6.471 &   6.463 &   6.474  &    6.451  &   6.473 &\\
\hline
MRE  & $-$0.06 \% & 0.24 \% & 0.44 \% & 0.29 \%   \\
MARE & 0.30 \% & 0.31 \% & 0.55 \% & 0.37 \%   \\
\end{tabular}
\end{ruledtabular}
\end{table*}

To evaluate the influence of the singles on the lattice constants
of covalent and metallic solid state systems, we show the equilibrium lattice constants for
selected materials in Tab. \ref{tab:lattice}.
We also use the present opportunity to evaluate whether
improved PAW potentials have an effect on the equilibrium lattice
constants. As shown in one of our recent work,\cite{klimes_NCPAW_2014}
quasiparticle energies as evaluated in the $GW$ approximation
can have large errors, since the  PAW projectors  possibly 
do not span the unoccupied orbital space accurately.
As a remedy to this problem, we have suggested to use PAW potentials
with norm-conserving
partial waves. These unfortunately increase the computational
cost, sometimes, even quite significantly. 
In Tab. \ref{tab:lattice}, the first column reports the lattice constants evaluated
using such norm-conserving PAW potentials. In the present
calculations, to attain the highest possible accuracy, the entire
lower lying core shell was included in the correlated calculations,
except for oxygen, carbon, nitrogen, fluorine (2$p$ elements) as well as phosphorus and chlorine. 
For instance for Si and Al, the 2$s$ and 2$p$ states were treated as valence states, for 
In and Sb the 4$d$, 4$s$ and 4$p$ states were fully included.
The calculations were performed using $6\times 6 \times 6$ k-points and
$8\times 8\times 8$ k-points
for gapped systems and metals, respectively.
Increasing the k-point set for selected semiconductors and insulators 
from $6\times 6 \times 6$ to  $8\times 8\times 8$ changed the lattice constants
by less than 0.1~\% (C, Si, SiC, LiCl). 
For metals, an increase of the k-point set to $10\times 10 \times 10$ k-points
changed the lattice constants also only by typically 0.1~\% (the results for
the transition metals and std-PAW are reported for $10\times 10 \times 10$ k-points).
This suggests that the lattice constants are k-point converged to about
0.1~\%. Errors incurred by the finite plane-wave basis set are of
the same order, so that we estimate the accuracy of the present
calculations to be about 0.2-0.3~\% in the lattice constants (or 1~\%
in the volumes).

For the RPA the mean relative error with respect to the zero-point
corrected experimental lattice constants is just 0.06~\% in the present calculations, 
whereas the mean absolute error is about 0.3~\%. We note that this is
within the estimated error bars of our calculations. It is therefore
futile to seek for any systematic errors: the RPA seems to be able
to predict lattice constants in almost perfect agreement with
experiment. Only for the Na metal, the lattice constant is obviously significantly underestimated
(excluding Na from the calculations, the MRE and MARE drop to 0.01~\% and 0.25~\% respectively).
There are very few density
functionals that yield a similar accuracy. In fact, the present
results slightly surpass those for the HSEsol
functional.\cite{Schimka_HSEsol_2011} The most commonly used functional, the PBE functional,
overestimates the lattice constants by about 1.3~\%, and
even the PBEsol functional yield a mean absolute relative error
of 0.46~\% for a slightly larger set.\cite{Schimka_HSEsol_2011} The use of not norm conserving PAW potentials
increases the lattice constants, on average by 0.3~\%. Also the mean absolute
relative error increases slightly from  0.3~\% to 0.35~\%.
For most elements, the differences between standard and NC PAW potentials are small. However, they can approach
up to 0.4~\% for elements with $3d$ and $4d$ semi-core states and (AlAs and GaAs)
and up to 1~\% for ionic compounds with strongly polarizable cores (Na).
Note that in ionic solids, vdW interactions are sizable, since the Na 2$s$ and 2$p$ core electrons 
are hardly screened and interact via van der Waals interactions with the neighboring halide atoms.

The origin for the increase in the lattice constants from the NC PAW potentials to the standard PAW potentials
is that the standard potentials  underestimate the polarizability of the core and this in turn yields too large
lattice constants.
For most applications, this small error of the lattice constants should be acceptable, however.
We finally note that the present values for the standard potentials are
also in good agreement with our first publication,\cite{harl2008}
although the potentials have been improved
since our previous calculations published in 2008. Specifically, the present
set of standard potentials preserves the norm of
the pseudo-orbitals better (although not perfectly, as the
NC PAW potentials do), which in turn increases the core polarizability and
decreases the lattice constants compared to the original 
values in Ref.~\onlinecite{harl2008}.

\begin{figure}
    \begin{center}
       \includegraphics[width=5.5cm,clip=true]{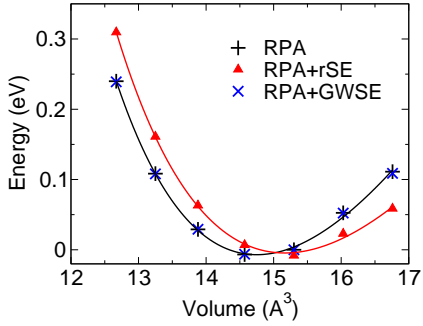}
    \end{center}
    \caption{
Energy-volume curves of Pd for RPA, RPA+rSE and RPA+GWSE. Clearly
the rSE increases the volume. Absolute energies are shifted
to yield zero at the minimum.
\label{fig:Pd}}

\end{figure}

Because the compute cost for the NC PAW potentials
is very high, we have evaluated the singles only for
standard potentials. The important result is that the singles
hardly change the lattice constants, except for some transition 
metals where an increase by 1 to 2~\% is observed  for
Cu and Pd. Fig. \ref{fig:Pd} shows that singles indeed shift significantly 
the equilibrium volume to larger values, 
clearly worsening the agreement with experiment. 
GWSE rectifies this, and yields almost identical
values to the RPA for all considered elements. 
As we will discuss in the next paragraph, the RPA density matrix for metals
is seemingly very different from the HF density matrix, and
most likely close to the DFT density matrix. This suggests
that HF singles are not adequate for metals, whereas, we
expect GW singles to be accurate across all systems.

\begin{figure}
    \begin{center}
       \includegraphics[width=6cm,clip=true]{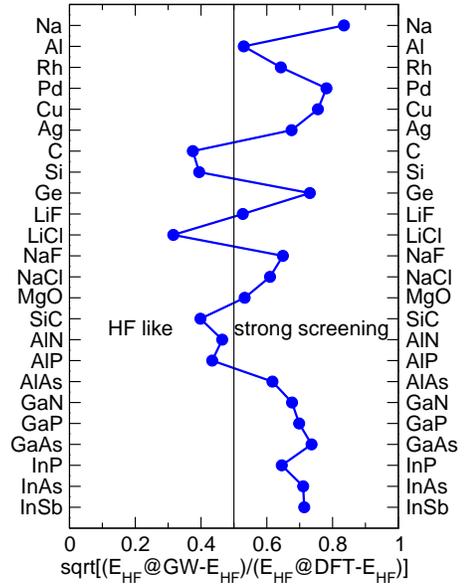}
    \end{center}
    \caption{
Measure of the ``distance'' between HF Green's function and 
RPA Green's function. See text for details. 
\label{fig:singles}}

\end{figure}

Fig. \ref{fig:singles} indicates how close the
GW density matrix is to the HF density matrix.
In order to measure this, one has to introduce a metric
to sensibly determine the difference. An obvious choice is
the total energy difference between the Hartree-Fock energy
evaluated using the HF density matrix  (rSE) and
the Hartree-Fock energy evaluated using the RPA/GW
density matrix
\[
 \Tr [\gamma_{\rm RPA}\hat h_{\rm HF} - \gamma_{\rm HF}\hat h_{\rm HF}].
\]
To present the values in a concise manner, we divide this
by 
\[
 \Tr [ \gamma_{\rm DFT}\hat h_{\rm HF} - \gamma_{\rm HF}\hat h_{\rm HF}],
\]
where $\gamma_{\rm DFT}$ is the DFT density matrix and finally  
take the square-root. The reason for including
the square-root is that the functional is variational and therefore quadratic 
around $\gamma_{\rm HF}$, since $\gamma_{\rm HF}$ is the groundstate
density matrix of $\hat h_{\rm HF}$. 
If the value is 0, the RPA density matrix coincides with the
HF density matrix, whereas for 1 it is closer to the DFT density matrix.
One clearly recognizes that the RPA density matrix is generally
quite close to the HF density matrix for light elements
and insulators. However, for metallic
materials Na, Pd, Cu, Rh,  and Ag, as well as  for small gap semiconductors
and semiconductors with heavy elements (Ge, Ga and In compounds)
this is not the case. The variational properties,  discussed above,
suggest that an evaluation of the mean field contribution
using HF orbitals will be generally superior to an evaluation
using DFT orbitals. In particular, for large gap systems, such
as rare gas solids, ice, but also C, Si and SiC, LiF and MgO,  
the differences between rSE and GWSE mean field contributions are tiny and only
of the order of 10~\% (essentially the square of the distance
shown in Fig.  \ref{fig:singles}).
For metals, this is, however, clearly not the case, and 
already shown in the previous paragraph, erroneously increases
the lattice constants compared to  RPA or RPA+GWSE.


\section{Discussion and Conclusions}

The present work is devoted to the performance of the random phase
approximation for extended systems if singles contributions are
taken into account. 
The first part of the paper focuses on the derivation of
the singles within the adiabatic-connection fluctuation-dissipation framework.
Not unexpectedly, this derivation yields exactly the same contributions as
the singles originally suggested by Ren and coworkers,\cite{ren2011}
because standard  Rayleigh-Schr\"odinger perturbation theory and
coupling-constant integration are identical. The coupling-constant integration and
the formulation used here has, however, the advantage 
that it gives a very clear picture of what the singles
describe. They account for the ``mean field''  energy changes
from the non-interacting Kohn-Sham reference system to the interacting
system, where we define mean field as those contributions arising
from a changes of the one-electron density matrix. 
This makes it very clear why cohesive energies are increased when
going from the DFT mean field description to the HF mean field description:
the latter contracts the orbitals and thus reduces the Pauli repulsion between 
the atoms or molecules.

Renormalized singles can be also derived in the present
framework and the final equation for them is particularly 
revealing (compare Equ. \eqref{equ:rSE}). The ``renormalized" singles describe the
energy difference between the Hartree-Fock eigenvalues and the diagonal
of the HF matrix evaluated using DFT orbitals (compare Equ. \eqref{equ:rSE_sum}).
This is not exactly identical to the renormalized singles suggested
by Ren et al. \cite{ren_renormalized_2013} although we found, in 
practice, that our simpler equation gives virtually the same
results as Ren's renormalized singles and, as a bonus, it  is trivial to implement
and most likely already available in most codes.

Inspired by the simple physically transparent form of the singles,
we have also suggested an alternative form for the singles that
relies on the RPA-density matrix instead of the HF-density matrix
(compare Equ. \eqref{equ:GWSE}). We have termed this correction GWSE,
singles in the $GW$ approximation. Clearly this description should be
superior to the simple HF description, as the RPA density matrix
should be fairly close to the exact groundstate one-particle density matrix.
In practice, for large band gap systems, such as rare gas solids, 
ice and adsorption of water on NaCl, differences between the
rSE and GWSE are small. This suggests that the HF density matrix
is often astonishingly close to the RPA
density matrix for systems with light atoms and large band gaps  (compare Fig. \ref{fig:singles}). In such cases, the rSE can be used instead of
the GWSE with little loss of accuracy. We believe this explains why the
rSE approximation was so successful for molecules. 
For metals and heavier elements, the approximation becomes increasingly
worse and an erroneousness increase of the lattice constants is
observed for some transition metals for the rSE approximation,
which is rectified by the  GWSE. 

We finally note that we have used this paper to revise our lattice
constants for main group elements and some transition metals
using highly accurate norm-conserving potentials. Although, the differences
to the previously published values are usually small and only of the
order of 0.3-0.5~\%, the present values should
be used as future reference.

\begin{acknowledgments}
This work was supported by the Austrian Science Fund (FWF) within the 
\emph{Spezialforschungsbereich Vienna Computational Materials Laboratory} (SFB
ViCoM, F41) and the \emph{Deutsche Forschungsgruppe Research Unit} FOR 1346. 
Computational resources were provided by the VSC (Vienna Scientifc Cluster) and the MetaCentrum under the program 
LM2010005 and the CERIT-SC under the program Centre CERIT Scientific Cloud, 
part of the Operational Program Research and Development for Innovations, 
Reg. no. CZ.1.05/3.2.00/08.0144.
\end{acknowledgments}

\appendix

\section{Fluctuation-dissipation expression}

Here we briefly derive the expression for the correlation energy, Equ.~(\ref{equ:acfdt})
starting from Equ. (\ref{eq:ccintegral_rhoc}).
The integrand in Equ. ~(\ref{eq:ccintegral_rhoc}) can be also written as
\begin{equation}
E_\lambda={1\over2}\int  {n_{2,\lambda}({\bf r}_1,{\bf r}_2)\over |{\bf r}_1-{\bf r}_2|} d^3 {\bf r}_1  d^3 {\bf r}_2\, ,
\end{equation}
where the two particle pair density is defined as 
\begin{equation}
n_{2,\lambda}({\bf r}_1,{\bf r}_2)=\langle \Psi_\lambda | \psi^\dagger({\bf r}_2)\psi^\dagger({\bf r}_1)
\psi({\bf r}_1)\psi({\bf r}_2)| \Psi_\lambda \rangle\,,
\end{equation}
and $\psi$ and $\psi^\dagger$ are the Fermionic annihilation and creation operators.
Using the common Fermionic anti-commutator relations one obtains:
\begin{equation}
\begin{split}
\label{eq_n2}
n_{2,\lambda}({\bf r}_1,{\bf r}_2)&=\langle \Psi_\lambda | \psi^\dagger({\bf r}_2)\psi({\bf r}_2) \psi^\dagger({\bf r}_1)\psi({\bf r}_1)| 
\Psi_\lambda \rangle\\
&-\delta({\bf r}_1-{\bf r}_2)
\underbrace{\langle \Psi_\lambda|\psi^\dagger({\bf r}_2) \psi({\bf
r}_1)|\Psi_\lambda \rangle }_{n_{1,\lambda}({\bf r}_1;{\bf r}_2)}\,.
\end{split}
\end{equation}
The response function is defined by Eq.~(\ref{eq:chi_def}), with the  density fluctuation operator in 
second quantization  given by:
\begin{equation}
\delta \hat n_\lambda(\tau,{\bf r}_1)=\psi^\dagger({\bf r}_1,\tau)\psi({\bf r}_1,\tau)-n_\lambda({\bf r}_1)\,.
\end{equation}
Using this expressions for both coordinates and simplifying the expression,
the response function takes the form:
\begin{widetext}
\begin{equation}
\chi_\lambda(\tau, {\bf r}_1, {\bf r}_2) =  - \langle \Psi_\lambda | \mathcal{T} [\psi^\dagger({\bf r}_1,\tau)\psi({\bf r}_1,\tau)  
 \psi^\dagger({\bf r}_2,0)\psi({\bf r}_2,0) ] | \Psi_\lambda \rangle + n_\lambda({\bf r}_1)n_\lambda({\bf r}_2)\,.
\end{equation}
We are interested in $\tau\rightarrow 0^-$, implying  the ${\bf r}_1$ operators act first;
after time ordering, we obtain (as $\chi_\lambda$ is bosonic, no sign change):
\begin{equation}
\chi_\lambda(0^-, {\bf r}_1, {\bf r}_2) =  - \langle \Psi_\lambda | \psi^\dagger({\bf r}_2,0)\psi({\bf r}_2,0)  
 \psi^\dagger({\bf r}_1, 0^-)\psi({\bf r}_1, 0^-)  | \Psi_\lambda \rangle + n_\lambda({\bf r}_1)n_\lambda({\bf r}_2)\,.
\end{equation}
The first term on the r.h.s. now equals the first term on the r.h.s. of Equ.~(\ref{eq_n2}).
By substituting for this term in Eq.~(\ref{eq_n2}) we obtain
\begin{equation}
n_{2,\lambda}({\bf r}_1,{\bf r}_2)=-\chi_\lambda(0^-, {\bf r}_1, {\bf r}_2)+ n_\lambda({\bf r}_1)n_\lambda({\bf r}_2)-\delta({\bf r}_1-{\bf r}_2)n_{1,\lambda}({\bf r}_1;{\bf r}_2)\,.
\end{equation}
\end{widetext}
This corresponds to  the terms depending on $\lambda$ in Eq.~(\ref{equ:acfdt}).
The terms not depending on $\lambda$ are obtained by  analogously rewriting $\langle \Psi_0 | \hat{v} |  \Psi_0\rangle$
in Equ. (\ref{eq:ccintegral_rhoc}).


%
\end{document}